\documentclass[final,5p,times,twocolumn]{elsarticle}

\usepackage{graphicx}
\usepackage{amssymb}
\usepackage{amsmath}
\usepackage{enumitem}
\usepackage{multirow}
\usepackage{tikz}
\usepackage{epstopdf}
\usepackage[percent]{overpic}
\usepackage{scrextend}
\usepackage{xcolor,xspace}
\usepackage[ulem=normalem]{changes}
\usepackage{hyperref}
\hypersetup{colorlinks=True,urlcolor=blue,linkcolor=blue,citecolor=blue,filecolor=black}
\usepackage{dcolumn,color,footnote,bm,braket}
\usepackage{url,longtable,tabularx}
\usepackage{threeparttable}
\usepackage{fancybox}

\usetikzlibrary{matrix}

\colorlet{Changes@Color}{red}

\newcommand\+{\dagger}
\newcommand\nb{n_{\rm b}}
\newcommand\hb{\hat H_{\rm B}}
\newcommand\vbf{\hat V_{\rm BF}}
\newcommand\bbt{\bar\beta}
\newcommand\bgam{\bar\gamma}

\newcommand\eb{E_{\rm B}(\beta;\xi)}
\newcommand\lba{\lambda_{K}(\beta;\eta)}
\newcommand\lbb{\lambda_{K}(\beta;\eta')}
\newcommand\cb{C_{\rm b}}
\newcommand\cbf{C_{\rm bf}}

\journal{Physics Letters B}

\begin{document}

\begin{frontmatter}

\title{Microscopic determination of
the interacting boson-fermion model Hamiltonian
from the nuclear energy density functional}

\author[1]{M. Homma}

\author[1,2]{K. Nomura\corref{cor}}
\ead{nomura@sci.hokudai.ac.jp}

\affiliation[1]
{
organization={Department of Physics, Hokkaido University},
city={Sapporo},
postcode={060-0810},
country={Japan}
}

\affiliation[2]
{
organization={Nuclear Reaction Data Center, Hokkaido University},
city={Sapporo},
postcode={060-0810},
country={Japan}
}

\cortext[cor]{Corresponding author}

\begin{abstract}
A microscopic formulation of the
interacting boson-fermion model
for odd-$A$ nuclei is made using
the nuclear energy density functional framework.
Strength parameters for
the bosonic Hamiltonian and
boson-fermion interactions
are shown to be determined completely
so that energy surfaces and
deformed single-particle energies
of the Bose-Fermi systems
should match the corresponding
self-consistent mean-field solutions
for fermionic systems.
In an illustrative application
to axially symmetric odd-$A$ Eu,
this procedure is shown to be valid
in describing
spherical-to-deformed
shape phase transitions in odd-$A$ and
even-even systems.
\end{abstract}

\date{\today}


\end{frontmatter}

Detailed and systematic
descriptions of odd-mass nuclei
have been a major challenge for
nuclear theory.
This is due to the presence
of an odd particle,
and one should treat
single-particle and collective
motions on the same footing
\cite{BM,RS}.
A viable approach is provided by
the interacting boson-fermion model (IBFM)
\cite{IBFM}, in which an even-even
nucleus is described
in terms of bosons that represent
nucleon pairs,
and is as a core
coupled to an unpaired nucleon.
The model has been successful for studying
low-lying states in odd-$A$ nuclei
\cite{IBFM}, identifying
nuclear supersymmetry
\cite{iachello1980susy,balantekin1981,frank2009},
and calculating
$\beta$
\cite{DELLAGIACOMA1989,nomura2022beta-ge}
and $\beta\beta$
\cite{yoshida2013,nomura2022bb}
decay rates.
While the IBFM has been applied
mostly on phenomenological grounds,
it should be connected
to a more fundamental
nuclear structure theory.

The interacting boson model (IBM)
\cite{IBM} for even-even nuclei
has certain microscopic foundations
on the underlying nucleonic dynamics
\cite{OAI,mizusaki1997,nomura2008}.
It was shown
\cite{nomura2008}, in particular,
that the IBM Hamiltonian
for general quadrupole states
\cite{nomura2010,nomura2011rot,nomura2012tri}
is determined
by means of the self-consistent
mean-field (SCMF) method
within the nuclear energy density
functional (EDF) framework
\cite{RS,bender2003,vretenar2005,robledo2019}.
The EDF-SCMF method also
produces spherical single-particle
energies and occupation probabilities,
which enter the boson-fermion
interactions in the generalized
seniority scheme \cite{scholten1985,IBFM}.
This reduces significantly
the number of parameters \cite{nomura2016odd},
but the boson-fermion
interaction strengths are, however,
fitted to observed low-energy levels
for each odd-$A$ nucleus.
The microscopic structure of the IBFM
in limited cases
has been pursued using schematic models
\cite{scholten1981boson,scholten1985,otsuka1987,yoshinaga2000}.
However, any unified and systematic
prescription to derive the
IBFM parameters from
a realistic nuclear structure
theory has not been reported.

To take a step further,
we develop a microscopic formulation
of the IBFM to completely determine
the corresponding strength parameters.
This procedure is also
based on the nuclear EDF,
but makes use of the
intrinsic-state framework of the IBFM
\cite{leviatan1988,leviatan1989,IBFM,iachello2011,petrellis2011}
producing energy surfaces and
deformed single-particle orbits,
which are made to match those in
the EDF-SCMF model.
The procedure is illustrated
in an application to odd-$A$ Eu isotopes,
which are, together with
the neighboring even-even Sm core nuclei,
empirically known to exhibit a quantum phase
transition (QPT) from nearly spherical
to prolate deformed shapes
\cite{iachello2001X5,casten2001X5,cejnar2010,iachello2011,petrellis2011}.
It is of interest to see if
the QPT-like features are realized
in the IBFM that is guided by the microscopic
framework of the EDF.

One might rather resort to a
fully fermionic and
consistent EDF approach to
odd-$A$ spectroscopy.
It was indeed done within the generator coordinate
method (GCM) with symmetry projections
and configuration
mixing of the EDF-SCMF solutions
\cite{RS,bender2003,robledo2019},
which explicitly takes into
account the breaking of
time-reversal symmetry and the blocking effects
\cite{bally2014,borrajo2016,zhou2024}.
The practical applications
of this approach are, however,
computationally highly demanding,
and have been limited to light
odd-$A$ (Mg) nuclei.
Thus, another aim of the present work
is to attain an alternative
EDF-based collective
model able to
predict spectroscopic properties
of heavy odd-$A$ nuclei including
those that are far from the
$\beta$-stability line.

The IBFM space is comprised of
the $s$ and $d$ bosons, which reflect collective
$J^{\pi}=0^+$ and $2^+$ pairs of valence nucleons
\cite{IBM,OAIT,OAI}, respectively,
and a single nucleon in orbits $j$.
Here we focus on single-$j$ systems.
The Hamiltonian $\hat H$
is written in general as
\begin{eqnarray}
\label{eq:ham}
 \hat H = \hat H_{\rm F} + \hb + \hat V_{\rm BF} \; .
\end{eqnarray}
The first term represents the single-particle
Hamiltonian
$\hat H_{\rm F} = \epsilon_j (a_j^\+\cdot\tilde a_j)\equiv \epsilon_j\hat n_j$,
with $\epsilon_j$, and $a^{(\+)}$ being the single-particle
energy and annihilation
(creation) operator, respectively.
We use the notation
$\tilde a_{j,m}=(-1)^{j-m}a_{j,-m}$,
with $m$ being the projection.
The second term stands for the
IBM Hamiltonian of the form
\begin{eqnarray}
\label{eq:bham}
 \hb=\epsilon_d \hat n_d + \kappa\hat Q\cdot\hat Q \; ,
\end{eqnarray}
where the first term
$\hat n_d = d^{\+}\cdot\tilde d$
represents the $d$-boson number operator,
with $\epsilon_d$ denoting the single-$d$
boson energy relative to that of $s$
bosons, and
the second term is the
quadrupole-quadrupole interaction
with the strength $\kappa$.
The quadrupole operator
$\hat Q=s^\+\tilde d + d^\+\tilde s + \chi(d^\+\times\tilde d)^{(2)}$,
where $\chi$ is a dimensionless parameter.
For the boson-fermion
interaction $\hat V_{\rm BF}$
in \eqref{eq:ham}, we take a form
with a minimal set of parameters,
but which is shown to be sufficient
for realistic calculations
\cite{IBFM,scholten1985}
\begin{align}
\label{eq:vbf}
\vbf = 
&\,
A \hat n_d\hat n_j 
+ \Gamma\hat Q\cdot(a_j^\+\times\tilde a_j)^{(2)}
\nonumber\\
&
+ \Lambda\sqrt{2j+1}:
\left[
(d^\+\times\tilde a_j)^{(j)}
\times
(a_j^\+\times\tilde d)^{(j)}
\right]^{(0)}: \; ,
\end{align}
where the first, second, and third terms are
referred to as the monopole,
quadrupole dynamical, and exchange terms,
respectively.
The quadrupole term represents
direct interactions, and the exchange term
reflects the fact that a boson is made of
a pair of fermions.
The monopole term has an effect of either
compress or stretch the whole energy spectrum.
The notation $:[\cdots]:$ in \eqref{eq:vbf}
denotes normal ordering,
and $A$, $\Gamma$, and $\Lambda$ are
the strength parameters.

To study the geometry of the IBFM Hamiltonian
the following basis is introduced:
\begin{eqnarray}
\label{eq:basis}
\ket{\nb;\bbt,\bgam}\otimes\ket{j,m} \; ,
\end{eqnarray}
which is a direct product of
the coherent state \cite{ginocchio1980}
of the $\nb$ interacting $s$ and $d$ bosons
\begin{eqnarray}
\label{eq:coherent}
 \ket{\nb;\bbt,\bgam}=(\nb!)^{-1/2}
(b_c^\+)^{\nb}\ket{0}
\end{eqnarray}
with
\begin{align}
 b_c^\+ = (1+\bbt^2)^{-1/2}
\left[
s^\+ + d_0^{\+}\bbt\cos{\bgam}
+\frac{1}{\sqrt{2}}(d_{+2}^\+ + d_{-2}^{\+})\bbt\sin{\bgam}
\right]
\end{align}
and the single-particle basis
$\ket{j,m}=a^{\+}_{j,m}\ket{0}$
for an odd fermion.
$\bbt$ and $\bgam$ in \eqref{eq:basis}
are boson analogs of the quadrupole
deformation $\beta$ and triaxiality $\gamma$,
respectively \cite{BM}, and
$\ket{0}$ represents the inert core.
The bosonic deformation $\bbt$ is related
to the fermionic counterpart $\beta$ so as to
be proportional to the latter
\cite{ginocchio1980,nomura2008}, i.e.,
\begin{eqnarray}
\label{eq:scale1}
 \bbt = \cb \beta \; ,
\end{eqnarray}
with $\cb$ being a constant of proportionality.
The relation \eqref{eq:scale1} takes
into account
the fact that the IBM model space is comprised
only of valence nucleons,
while in the geometrical model
all constituent nucleons are involved.
The bosonic $\bgam$ deformation is, however,
the same angle variable as the triaxiality
$\gamma$, i.e., $\bgam = \gamma$.
The expectation value
of the Hamiltonian \eqref{eq:ham} in the basis
\eqref{eq:basis} gives an energy surface
in a matrix form \cite{leviatan1988}.
As an initial study,
we assume axial symmetry
with $\gamma=0^{\circ}$,
in which case
the energy surface is put into
a diagonal matrix
with its elements corresponding to
values of the projection $m=K$:
\begin{eqnarray}
\label{eq:pes}
 E_{K}(\beta)=\eb + \lba \; ,
\end{eqnarray}
where $\eb$, with
$\xi$ denoting a set of the parameters
$\xi=\{\epsilon_d\,,\kappa\,,\chi\,,\cb\}$,
is the bosonic energy surface
obtained as the expectation
value $\bra{\nb;\bbt}{\hb}\ket{\nb;\bbt}$,
and is given by
\begin{align}
\label{eq:bpes}
 \eb
=&\,
\frac{\nb\bbt^2}{1+\bbt^2}
\left\{
5\kappa
+
\left[\epsilon_d+\kappa(1+\chi^2)\right]\bbt
\right\}
\nonumber\\
&
+\frac{\nb(\nb-1)\bbt^2}{(1+\bbt^2)^2}
\left(
2-\chi\sqrt{\frac{2}{7}}\bbt
\right)^2 \; .
\end{align}
The second term in \eqref{eq:pes}
represents the single-particle energy
dependent on the deformation $\beta$
\cite{leviatan1988,IBFM}:
\begin{align}
 \label{eq:nilsson}
&\lba
=\epsilon_j
+ A\frac{\nb\bbt^2}{1+\bbt^2}
+ \frac{\nb\bbt}{1+\bbt^2}
\left[3K^2-j(j+1)\right]P_j
\nonumber\\
&
\times
\left\{
\Gamma\left(\bbt\chi\sqrt{\frac{2}{7}}-2\right)
-\Lambda P_j\bbt\sqrt{2j+1}\left[3K^2-j(j+1)\right]
\right\} \; ,
\end{align}
with $\eta=\{\,A\,,\Gamma\,,\Lambda\,\}$
and $P_j=[(2j-1)j(2j+1)(j+1)(2j+3)]^{-1/2}$.
As it is conventionally made,
the parameter $\chi$ in \eqref{eq:nilsson}
is assumed to be the same as that
in \eqref{eq:bpes}.

The first step to determine
the IBFM parameters
is a set of the constrained SCMF calculations
of the potential energy surfaces
for even-even (Sm) core nuclei.
The constraints are on the mass quadrupole
moment $Q_{20}$, which is related to the
geometrical deformation $\beta$ \cite{BM}.
The constrained calculations are performed
by means of
the Hartree-Fock-Bogoliubov (HFB) approach using
the Skyrme SkM* force \cite{skyrme,skms}.
The computer program HFBTHO (v4.0) \cite{hfbtho400}
is used for the mean-field calculations
throughout this work.

Second, the parameters
for the boson-core Hamiltonian $\hb$ ($\xi$)
is determined \cite{nomura2008,nomura2010}
by mapping the one-dimensional
potential energy
curve (PEC) obtained by the constrained HFB
onto the IBM energy curve of \eqref{eq:bpes}.
These parameters are calibrated so that
basic characteristics of
the bosonic energy curve in the
vicinity of the equilibrium minimum,
that is, the depth of the potential,
and curvature up to a few MeVs
from the minimum, should match
those of the HFB PEC.

%
%
\begin{figure}[ht]
\begin{center}
\includegraphics[width=\linewidth]{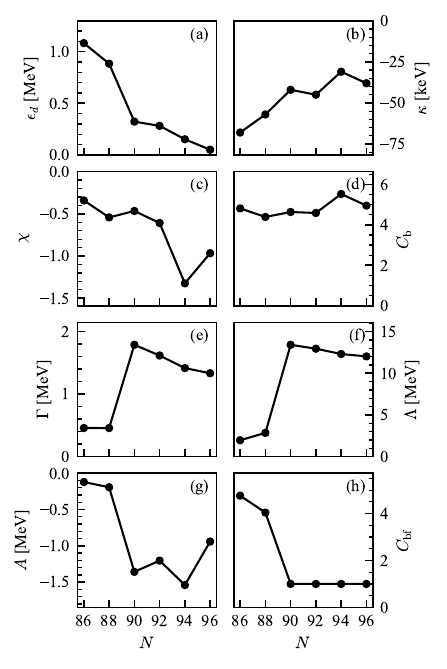}
\caption{
Derived values of the IBM parameters
for $^{148-158}$Sm [(a)--(d)],
and IBFM parameters for $^{149-159}$Eu
[(e)-(h)].
}
\label{fig:para}
\end{center}
\end{figure}

Figures~\ref{fig:para}(a)--\ref{fig:para}(d)
depict the derived IBM parameters.
With increasing neutron number $N$,
the $\epsilon_d$ decreases,
$\kappa$ gradually decreases in magnitude,
and $\chi$ decreases
to be a large negative value close
to the rotational SU(3) limit,
$\chi\approx -\sqrt{7}/2$.
The absolute values and behaviors with $N$
of these parameters are overall
consistent with earlier microscopic
\cite{scholten1978,nomura2008},
and phenomenological \cite{IBM}
IBM calculations,
and reflect the transition
from nearly spherical to deformed
shapes.
The derived value of the
scale factor $\cb$ is, however, rather
insensitive to $N$.
This factor is determined in the mapping
so that the location of the minimum
in the HFB PEC is reproduced.
Note that some
local irregularities found in
$\kappa$, $\chi$, and $\cb$ at $N\approx94$
are considered to be a consequence
of the numerical fit.

The final step is to fix the
interaction $\vbf$.
We propose to determine the
$A$, $\Gamma$, $\Lambda$
values so that the single-particle
energies of the IBFM, $\lba$
\eqref{eq:pes},
should reproduce the behaviors of the
HFB single-particle energies,
$\epsilon_K(\beta)$,
calculated for the odd-$A$ nuclei.
The PEC-mapping 
for the even-even Sm is adequately
made so as to reproduce the
HFB PEC typically in the domain
$0\leqslant\beta\leqslant\beta_{e}$,
with $\beta_e$ corresponding to
the equilibrium minimum.
Accordingly,
we associate $\lba$
with the HFB single-particle spectra
within the range from $\beta=0$ up to
the deformation corresponding
to the minimum in
the HFB PEC for odd-$A$ Eu.
Note that since
the equilibrium minimum for an
odd-$A$ Eu nucleus occurs at almost
the same amount of
deformation as $\beta_e$ for
the even-even Sm core,
we do not distinguish
the equilibrium deformation
of the odd-$A$ system from
that of the even-even system,
using the common notation $\beta_e$.
In particular, $\lba$
are made equal to the HFB
counterparts at
$\beta\approx\beta_e$:
\begin{eqnarray}
\label{eq:spe}
 \lambda_K(\beta_e;\eta) \approx \epsilon_K(\beta_e) \; ,
\end{eqnarray}
for each $K$.
We note that the region corresponding to
a much larger deformation $\beta\gg\beta_e$
is not of much relevance for our purpose.
For the very large quadrupole deformation
the IBFM single-particle
energies become flat, whereas
those from the self-consistent
calculations in general
exhibit much more significant changes
with $\beta$.
This is, to a large extent, due to
the limited IBFM model space,
consisting of valence nucleons.
The comparison of the IBFM with
the HFB single-particle orbits
in the region $\beta\gg\beta_e$
would, therefore, not make much sense.

The relation \eqref{eq:spe} holds
to a good extent
for those odd-$A$ nuclei with $N\leqslant88$,
in which the deformation is relatively
weak ($\beta_e \approx 0.15$),
and the HFB and IBFM
single-particle spectra do not differ
significantly in the region $\beta\lesssim\beta_e$.
However, for those nuclei with $N\geqslant90$,
which exhibit a larger
equilibrium deformation $\beta_e\gtrsim0.2$,
it turns out that the IBFM single-particle
levels differ significantly from
the HFB ones at $\beta\approx\beta_e$,
the reason being, as mentioned above,
the limited IBFM configuration space
and analytic form of
$\lba$ \eqref{eq:nilsson}.
To make a meaningful comparison with
the HFB single-particle energies
for these deformed nuclei,
we assume that the scale
factor $\cb$ \eqref{eq:scale1}
involved in
$\lba$ \eqref{eq:nilsson}
is allowed to differ
from that for the bosonic
energy surface $\eb$ \eqref{eq:bpes}.
This is led by the fact that
this factor is the most relevant
among those parameters
involved in \eqref{eq:nilsson}
that accounts for the difference between
the IBFM and HFB spaces.
In carrying out
the mapping of $\lba$
for these deformed nuclei, therefore,
$\cb$ in \eqref{eq:nilsson}
is now treated as an additional
fitting parameter,
and is hereafter referred to as $\cbf$,
in order to distinguish from $\cb$.
Rewriting $\lba$
as $\lbb$,
with $\eta'=\{\,A\,,\Gamma\,,\Lambda\,,\cbf\,\}$,
a generalized formula
\begin{eqnarray}
\label{eq:pes1}
 E_{K}(\beta)=\eb + \lbb
\end{eqnarray}
is hereafter considered.
The different scale factors
are chosen for $\eb$ and $\lbb$
in deformed region,
in order that both the PEC for the
even-even core and deformed
single-particle levels for
the odd-$A$ neighbor
obtained from the HFB calculations
should be reproduced.
Also the criterion for
using the scale factor $\cbf\neq\cb$
for $^{153-159}$Eu, but $\cbf=\cb$
for $^{149}$Eu and $^{151}$Eu,
is based on
the empirical fact that
the former nuclei correspond to
the regions at the critical point
of and {\it after} the QPT,
in which a strong quadrupole
deformation occurs,
and the latter nuclei correspond
to the region {\it before}
the QPT, characterized by a
moderate deformation.

%
%
\begin{figure}[ht]
\begin{center}
\includegraphics[width=\linewidth]{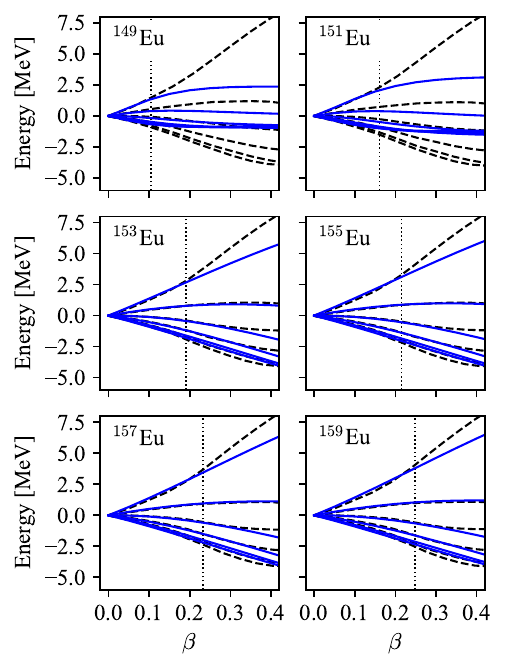}
\caption{
Deformed single-particle spectra
originating from the proton $1h_{11/2}$
spherical orbit
for the odd-$A$ nuclei $^{149-159}$Eu,
calculated by the Skyrme HFB
[$\epsilon_K(\beta)$, dashed lines]
and IBFM [$\lbb$, solid lines].
The vertical dotted line indicates
$\beta_e$ in the
HFB PEC for Sm/Eu.
For each nucleus,
different IBFM single-particle spectra
correspond to, in the
increasing order in energy,
$K={1/2}$, ${3/2}$, ${5/2}$,
${7/2}$, ${9/2}$, and ${11/2}$,
and should be compared with
the HFB counterparts, which, 
using the asymptotic quantum
numbers $K[Nn_zm_l]$,
stand for, in the
increasing order in energy,
the
${1/2}[550]$, ${3/2}[541]$,
${5/2}[532]$, ${7/2}[523]$,
${9/2}[514]$, and
${11/2}[505]$ orbits, respectively.
}
\label{fig:nilsson}
\end{center}
\end{figure}

Figure~\ref{fig:nilsson} shows
the deformed single-particle
energies for the
odd-$A$ nuclei $^{149-159}$Eu
originating from the proton
spherical orbit $1h_{11/2}$,
which are obtained from the constrained
HFB calculations (dashed lines).
For these HFB calculations,
blocking effects are taken into account
at each $\beta$,
and for all possible single-particle orbits.
The IBFM single-particle
energies $\lbb$, also
depicted in Fig.~\ref{fig:nilsson}
(solid lines),
exhibit in the range
$0\leqslant\beta\leqslant\beta_e$
behaviors that are consistent with
those of the HFB counterparts
for most of the Eu nuclei.

The derived parameters
for the odd-$A$ $^{149-159}$Eu
are shown in
Figs.~\ref{fig:para}(e)--\ref{fig:para}(h).
A drastic change in magnitude
of most of these parameters
from $N=88$ to 90 reflects
the nuclear structure change.
The present $\Gamma$ values are
substantially larger,
and exhibit a more rapid change
with $N$ than those in earlier
IBFM calculations
\cite{scholten1985,nomura2016odd,nomura2017odd-2}.
The abrupt increase
of $\Lambda$ at $N=88$ or 90
is consistent with
those in the previous IBFM
calculations for Eu
using the generalized seniority
and the relativistic \cite{nomura2016odd}
and nonrelativistic (Gogny) EDFs
\cite{nomura2017odd-2} as inputs.
The present $\Lambda$ values are
larger than these values
by a factor of 2 for deformed nuclei
with $N\geqslant90$.
The monopole strength $A$ exhibits
an increase in magnitude with $N$,
while the previous values
\cite{nomura2016odd,nomura2017odd-2}
showed an opposite trend.
Of particular interest is the fact that
the derived $\cbf$ for deformed nuclei with
$N\geqslant 90$ is approximately equal to unity,
meaning that the IBFM
single-particle energy is dictated by
the same amount of deformation
as the geometrical one.

%
%
\begin{figure}[ht]
\begin{center}
\includegraphics[width=\linewidth]{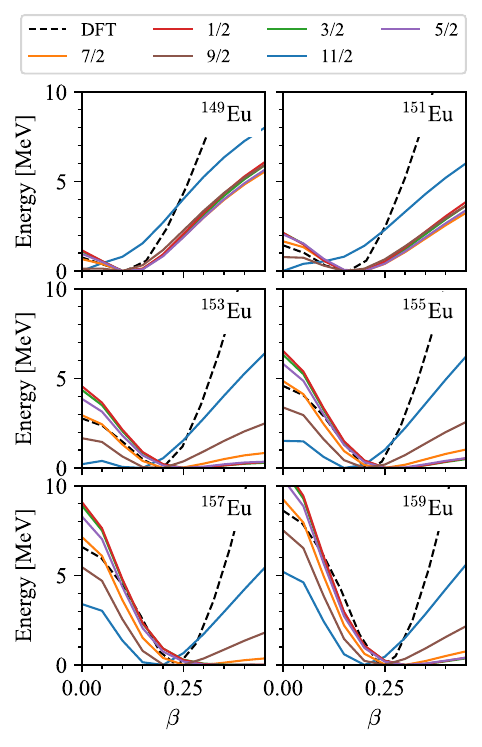}
\caption{
Potential energy curves corresponding to
$K=1/2$, $3/2$, $5/2$, $7/2$, $9/2$, and $11/2$
for $^{149-159}$Eu resulting from the
IBFM [$\lbb$, solid lines],
and the Skyrme-HFB energy
curve (dashed lines).
Each energy curve is shown with respect
to the minimum.
}
\label{fig:pes-eu}
\end{center}
\end{figure}

Figure~\ref{fig:pes-eu} displays
the IBFM PECs \eqref{eq:pes1}
for $^{149-159}$Eu, corresponding to
$K=1/2$, $3/2$, $5/2$, $7/2$, $9/2$, and $11/2$
using the derived parameters shown
in Fig.~\ref{fig:para}.
The Skyrme-HFB PEC, also shown in
Fig.~\ref{fig:pes-eu},
exhibits a weakly deformed
minimum with the depth of
$\approx1$ MeV for $^{149}$Eu and $^{151}$Eu.
The potential becomes even more sharper
from $^{153}$Eu to $^{155}$Eu,
and to $^{157}$Eu.
The equilibrium deformation
also jumps from $\beta_e\approx0.15$
to $0.20$ at $N=90$, and keeps
increasing to be $\beta_e\approx0.25$ 
for $^{157}$Eu and $^{159}$Eu.
As there are six IBFM energy curves
for $j={11/2}$,
a direct comparison with the
HFB energy curve is not straightforward.
But in Fig.~\ref{fig:pes-eu},
all these different IBFM PECs
show qualitatively similar systematic
to the HFB counterpart,
that is, the potential becomes
deeper with $N$, and the minimum
shifts rather abruptly at
$N\approx90$.
We can thus infer from the behaviors
of the energy curves of both the
IBFM and HFB that a phase
transitional, abrupt nuclear structure
change occurs in the odd-$A$ Eu nuclei
at the mean-field level.

%
%
\begin{figure*}[ht]
\begin{center}
\includegraphics[width=\linewidth]{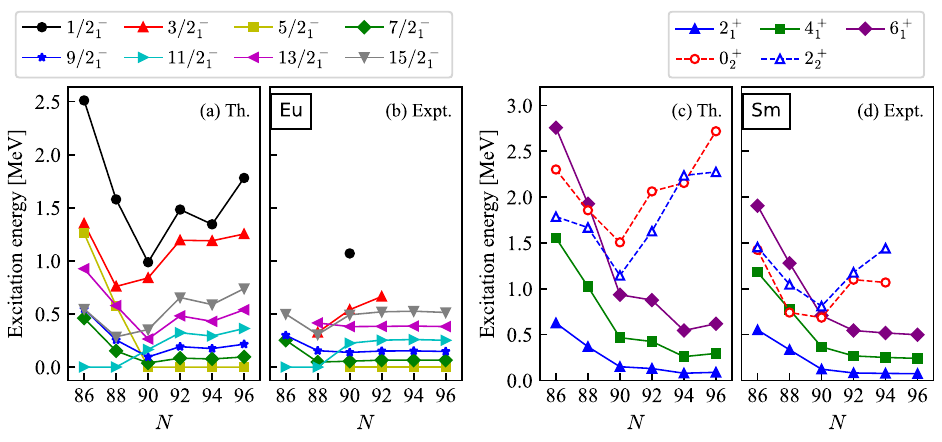}
\caption{
Calculated and experimental \cite{data} low-energy
spectra for odd-$A$ $^{149-159}$Eu [(a) and (b)]
and even-even $^{148-158}$Sm [(c) and (d)].
}
\label{fig:level}
\end{center}
\end{figure*}

We now turn to low-lying states.
Figure~\ref{fig:level}(a) shows the
calculated
energy spectra of negative-parity
yrast states in the odd-$A$
nuclei $^{149-159}$Eu.
The excitation energies are eigenvalues of
the mapped IBFM with derived parameters shown
in Fig.~\ref{fig:para}, resulting from the
numerical diagonalization in the basis
$\ket{L,j\,;I}$ \cite{NPBOS},
with $L$, and $I$ being the boson,
and total angular momenta,
respectively.
A signature of the QPT
is a change in the spin of the
lowest-energy state of a given parity
at a particular nucleon number
\cite{petrellis2011}.
In Figs.~\ref{fig:level}(a)
and \ref{fig:level}(b),
the calculation suggests that
the ground-state spin changes from
$I={11/2}^-$ to ${5/2}^-$ at $N=90$,
which is consistent with experiment.
The ${11/2}^-$ ground-state
for $N=86$ and $88$ is interpreted
to be made of the configurations of
the odd fermion at $1h_{11/2}$
orbit that is weakly
coupled to the moderately deformed
even-even boson core.
The ground state band based on
the ${11/2}^-$ state follows the $\Delta I=2$
level patterns of the weak-coupling limit.
For deformed nuclei with $N>90$,
the yrast levels form a rotational-like
band built on the ${5/2}^-$ ground state,
which exhibits a $\Delta I=1$ sequence
of levels characteristic
of the strong-coupling picture.
The IBFM reasonably
reproduces these level structures,
while it overestimates the
observed ground-state bands
for those nuclei with $N\geqslant92$
and ${3/2}^-$ energies at
$N=88$, 90, and 92.

We also show
in Figs.~\ref{fig:level}(c)
and \ref{fig:level}(d)
the excitation energies
for the even-even core nuclei $^{148-158}$Sm.
The mapped IBM reproduces
fairly well
overall patterns of the observed \cite{data}
energy levels.
The yrast levels
exhibit a decrease with $N$ both
in theory and experiment.
The decrease of the ratio of
the $4^+_1$ to $2^+_1$
energies $R_{4/2}$ is an indicator
of a shape phase transition from
the vibrational U(5)
(with $R_{4/2}\approx 2.0$)
to rotational SU(3)
(with $R_{4/2}\approx 3.3$)
limits.
This ratio is calculated to
be 2.49, 2.79, 3.13, 3.29, 3.32, and 3.34
for $^{148-158}$Sm, respectively,
which are consistent with the
experimental values 2.15, 2.32, 3.00,
3.30, 3.29, and 3.29.
The overestimates of $R_{4/2}$
for $^{148}$Sm and $^{150}$Sm
reflect that the mapped IBM
gives rather deformed rotational
structure, the reason being that the
HFB PEC exhibits a very
sharp potential \cite{nomura2010}.
The qualitative features of
the calculated $0^+_2$ and $2^+_2$
levels reaching minima in energy
at the transitional nucleus
$^{152}$Sm ($N=90$) are consistent with
experiment.
The calculation overestimates
the $0^+_2$ and $2^+_2$
energy levels,
which could also be attributed to
the properties of the HFB PECs.
It should be, however, noted that
the $0^+_2$ states in rare-earth region
may be of character that cannot
be accounted for in the standard IBM,
e.g., intruder excitations.
An extension of the IBFM to
include the intruder states and
configuration mixing was recently
made for odd-$A$ Nb nuclei
\cite{gavrielov2022,leviatan2025,MAYABARBECHO2025}.
Implementations of these effects
in the mapping procedure will be
an interesting future work.

In summary, we have presented
a microscopic determination of the
IBFM strength parameters
from the EDF-SCMF calculations,
in which the energy surfaces and
deformed single-particle levels of
the IBFM are made to match
the HFB counterparts.
An illustrative application
to axially symmetric odd-$A$ Eu nuclei
revealed the validity of the method in
describing consistently a QPT-like transition
from the nearly spherical
to prolate deformed shapes
in odd-$A$ and even-even systems.
The method can be extrapolated to
predict spectroscopic properties of
odd-$A$ exotic nuclei in the regions
for which experimental data do not exist.
This opens up new possibilities of
studying those nuclear properties
that are of broader physical significance,
including the shape coexistence
in neutron-rich odd-$A$ nuclei,
and fundamental processes such as
$\beta$ and $\beta\beta$ decays.

\section*{Acknowledgements}

This work has been supported
by JST SPRING Grant No. JPMJSP2119, and
by JSPS KAKENHI Grant No. JP25K07293.


\bibliographystyle{elsarticle-num} 
\bibliography{ref}

\begin{thebibliography}{10}
\expandafter\ifx\csname url\endcsname\relax
  \def\url#1{\texttt{#1}}\fi
\expandafter\ifx\csname urlprefix\endcsname\relax\def\urlprefix{URL }\fi
\expandafter\ifx\csname href\endcsname\relax
  \def\href#1#2{#2} \def\path#1{#1}\fi

\bibitem{BM}
A.~Bohr, B.~R. Mottelson, Nuclear Structure, Benjamin, New York, 1975.

\bibitem{RS}
P.~Ring, P.~Schuck, The nuclear many-body problem, Springer, Berlin, 1980.

\bibitem{IBFM}
F.~Iachello, P.~{Van Isacker}, The interacting boson-fermion model, Cambridge
  University Press, Cambridge, 1991.

\bibitem{iachello1980susy}
F.~Iachello, Phys. Rev. Lett. 44 (1980) 772.

\bibitem{balantekin1981}
A.~B. Balantekin, I.~Bars, F.~Iachello, Nucl. Phys. A 370 (1981) 284.

\bibitem{frank2009}
A.~Frank, J.~Jolie, P.~{Van Isacker}, Symmetries in atomic nuclei, Springer,
  2009.

\bibitem{DELLAGIACOMA1989}
F.~Dellagiacoma, F.~Iachello, Phys. Lett. B 218 (1989) 399.

\bibitem{nomura2022beta-ge}
K.~Nomura, Phys. Rev. C 105 (2022) 044306.

\bibitem{yoshida2013}
N.~Yoshida, F.~Iachello, Prog. Theor. Exp. Phys. 2013 (2013) 043D01.

\bibitem{nomura2022bb}
K.~Nomura, Phys. Rev. C 105 (2022) 044301.

\bibitem{IBM}
F.~Iachello, A.~Arima, The interacting boson model, Cambridge University Press,
  Cambridge, 1987.

\bibitem{OAI}
T.~Otsuka, A.~Arima, F.~Iachello, Nucl. Phys. A 309 (1978) 1.

\bibitem{mizusaki1997}
T.~Mizusaki, T.~Otsuka, Prog. Theor. Phys. Suppl. 125 (1996) 97.

\bibitem{nomura2008}
K.~Nomura, N.~Shimizu, T.~Otsuka, Phys. Rev. Lett. 101 (2008) 142501.

\bibitem{nomura2010}
K.~Nomura, N.~Shimizu, T.~Otsuka, Phys. Rev. C 81 (2010) 044307.

\bibitem{nomura2011rot}
K.~Nomura, T.~Otsuka, N.~Shimizu, L.~Guo, Phys. Rev. C 83 (2011) 041302.

\bibitem{nomura2012tri}
K.~Nomura, N.~Shimizu, D.~Vretenar, T.~Nik{\v{s}}i\'c, T.~Otsuka, Phys. Rev.
  Lett. 108 (2012) 132501.

\bibitem{bender2003}
M.~Bender, P.-H. Heenen, P.-G. Reinhard, Rev. Mod. Phys. 75 (2003) 121.

\bibitem{vretenar2005}
D.~Vretenar, A.~V. Afanasjev, G.~A. Lalazissis, P.~Ring, Phys. Rep. 409 (2005)
  101.

\bibitem{robledo2019}
L.~M. Robledo, T.~R. Rodríguez, R.~R. Rodríguez-Guzmán, J. Phys. G: Nucl.
  Part. Phys. 46 (2019) 013001.

\bibitem{scholten1985}
O.~Scholten, Prog. Part. Nucl. Phys. 14 (1985) 189.

\bibitem{nomura2016odd}
K.~Nomura, T.~Nik{\v{s}}i\'c, D.~Vretenar, Phys. Rev. C (2016) 054305.

\bibitem{scholten1981boson}
O.~Scholten, A.~Dieperink, On the boson-fermion interaction, in: F.~Iachello
  (Ed.), Interacting Bose-Fermi Systems in Nuclei, Springer, 1981, pp.
  343--353.

\bibitem{otsuka1987}
T.~Otsuka, N.~Yoshida, P.~Van~Isacker, A.~Arima, O.~Scholten, Phys. Rev. C 35
  (1987) 328.

\bibitem{yoshinaga2000}
N.~Yoshinaga, Y.~D. Devi, A.~Arima, Phys. Rev. C 62 (2000) 024309.

\bibitem{leviatan1988}
A.~Leviatan, Phys. Lett. B 209 (1988) 415.

\bibitem{leviatan1989}
A.~Leviatan, B.~Shao, Phys. Rev. Lett. 63 (1989) 2204.

\bibitem{iachello2011}
F.~Iachello, A.~Leviatan, D.~Petrellis, Phys. Lett. B 705 (2011) 379.

\bibitem{petrellis2011}
D.~Petrellis, A.~Leviatan, F.~Iachello, Ann. Phys. 326 (2011) 926.

\bibitem{iachello2001X5}
F.~Iachello, Phys. Rev. Lett. 87 (2001) 052502.

\bibitem{casten2001X5}
R.~F. Casten, N.~V. Zamfir, Phys. Rev. Lett. 87 (2001) 052503.

\bibitem{cejnar2010}
P.~Cejnar, J.~Jolie, R.~F. Casten, Rev. Mod. Phys. 82 (2010) 2155.

\bibitem{bally2014}
B.~Bally, B.~Avez, M.~Bender, P.-H. Heenen, Phys. Rev. Lett. 113 (2014) 162501.

\bibitem{borrajo2016}
M.~Borrajo, J.~L. Egido, Eur. Phys. J. A 52 (2016) 277.

\bibitem{zhou2024}
E.~F. Zhou, X.~Y. Wu, J.~M. Yao, Phys. Rev. C 109 (2024) 034305.

\bibitem{OAIT}
T.~Otsuka, A.~Arima, F.~Iachello, I.~Talmi, Phys. Lett. B 76 (1978) 139.

\bibitem{ginocchio1980}
J.~N. Ginocchio, M.~W. Kirson, Nucl. Phys. A 350 (1980) 31.

\bibitem{skyrme}
T.~H.~R. Skyrme, Nucl. Phys. 9 (1958) 615.

\bibitem{skms}
J.~Bartel, P.~Quentin, M.~Brack, C.~Guet, H.-B. Hakansson, Nucl. Phys. A 386
  (1982) 79.

\bibitem{hfbtho400}
P.~Marevi\'c, N.~Schunck, E.~Ney, R.~{Navarro P\'erez}, M.~Verriere, J.~O'Neal,
  Comput. Phys. Commun. 276 (2022) 108367.

\bibitem{scholten1978}
O.~Scholten, F.~Iachello, A.~Arima, Ann. Phys. 115 (1978) 325.

\bibitem{nomura2017odd-2}
K.~Nomura, R.~Rodr\'{\i}guez-Guzm\'an, L.~M. Robledo, Phys. Rev. C 96 (2017)
  014314.

\bibitem{data}
{Brookhaven National Nuclear Data Center}, {http://www.nndc.bnl.gov}.

\bibitem{NPBOS}
T.~Otsuka, N.~Yoshida, {}JAERI-M (Japan At. Ener. Res. Inst.) Report No. 85
  (1985).

\bibitem{gavrielov2022}
N.~Gavrielov, A.~Leviatan, F.~Iachello, Phys. Rev. C 106 (2022) L051304.

\bibitem{leviatan2025}
A.~Leviatan, N.~Gavrielov, Phys. Lett. B 868 (2025) 139647.

\bibitem{MAYABARBECHO2025}
E.~Maya-Barbecho, J.-E. García-Ramos, Phys. Lett. B 868 (2025) 139724.

\end{thebibliography}

\end{document}